\documentclass[pra,twocolumn,titlepage,nofootinbib,amsmath,showpacs]{revtex4}%
\usepackage{amsmath}
\usepackage{amsfonts}
\usepackage{amssymb}
\usepackage{graphicx}%
\begin{document}
\title{HFS interval of the $2s$ state of hydrogen-like atoms
and a constraint on a pseudovector boson with mass
below $1\;$keV$/c^2$}
\author{S.~G.~Karshenboim}
\email{savely.karshenboim@mpq.mpg.de} \affiliation{D.~I. Mendeleev
Institute for Metrology, St.Petersburg, 190005, Russia
\\ {\rm and}
Max-Planck-Institut f\"ur Quantenoptik, Garching, 85748, Germany}

\begin{abstract}
A constraint on a long-range spin-dependent interaction
$\alpha^{\prime\prime}({\bf s}_1\cdot{\bf s}_2)\,{\rm e}^{-\lambda
r}/r$, which can be induced by a pseudovector light boson, is
presented. We study theoretical and experimental data on a specific
difference $8\times E_{\rm hfs}(2s)-E_{\rm hfs}(1s)$ for light
two-body atoms. The spin-dependent coupling constant
$\alpha^{\prime\prime}$ of electron-nucleus interaction in hydrogen,
deuterium and helium-3 ion is constrained at the level below a
part in $10^{16}$. The derived constraints are related to the range
of masses below $4\;{\rm keV}/c^2$.
\end{abstract}
\maketitle

\section{Introduction}

A strong constraint from atomic physics can be set on a
spin-dependent long-range interaction induced by a light
axial-vector particle. In principle, a constraint on light particle
with mass in keV/$c^2$ range may be derived by many means, involving
cosmological estimation \cite{cosmo} and astrophysical phenomena
\cite{astro}. Such constraints involve a number of parameters, such
as the particle mass, its coupling to other particles, lifetime etc.
In contrast to that a constraint based on limiting a possible
deviation of electron-nucleus interaction in atomic distance range
depends on only two parameters, namely the particle mass $\lambda$
and a strength of interaction between an electron and a nucleus,
mediated by the intermediate particle under consideration.

The previous constraint of this kind on spin-dependent interaction
was derived from data on the hyperfine structure (HFS) interval of
the $1s$ state in light hydrogen-like atoms \cite{prl,prespin}. The
result was for a particle substantially lighter than $4\;{\rm
keV}/c^2$ and the accuracy was limited either by the HFS experiment
(muonium, positronium) or by an uncertainty of the related
nuclear-effect contribution (hydrogen, deuterium). (It has been also
extended there to heavier particles but with a reduced constraining
strength.)

Here, to avoid uncertainties due to nuclear effects, we consider a
specific difference of the $1s$ and $2s$ hyperfine intervals
\begin{equation}
D_{21}=8\times E_{\rm hfs}(2s)-E_{\rm hfs}(1s)\;,
\end{equation}
which is essentially free of such a problem \cite{d21th,my_rep}.
Experimental data with appropriate accuracy are available for
hydrogen \cite{exph1,exph2s}, deuterium \cite{expd1,expd2s} and
helium-3 ion \cite{exphe1,exphe2s} for their $1s$ and $2s$ hyperfine
intervals. The corresponding data are summarized in
Appendix~\ref{s:a1}. For theoretical results, which are summarized
in Appendix~\ref{s:a2}, we follow \cite{d21th1}.

Theory suggests that there is a massive cancelation of various
contributions, which are proportional to the squared value of the
wave function at origin
\[|\Psi_{ns}(0)|^2\propto n^{-3}\;.\]
Those
include various uncertain nuclear-effect terms, and a theoretical
prediction for the difference has a very safe ground and has reached
a high accuracy.

That is not the only theoretical advantage of the difference. The
cancelation happens also with the leading term (see below) and
because of that the fractional uncertainty of measurements of the
difference is relatively low. Even with such a fractional accuracy
the difference remains very sensitive to many higher-order effects.

Meantime, the theoretical accuracy in QED calculations for the HFS
intervals is strongly affected by accuracy of our knowledge of
fundamental constants required for the calculations and in
particular of the nuclear magnetic moments (see, e.g.
\cite{my_rep}). In the case the of difference the leading
contributions have a large theoretical uncertainty, however, they
cancel out in the difference and as result the theory of the
difference is relatively immune to any problems in determination of
the magnetic moments and other fundamental constants, which is
indeed quite advantageous for theoretical calculations.

Returning to the leading term, the cancelation happens for the
leading term to the $ns$ HFS interval, a so-called Fermi
contribution
\begin{equation}\label{ef}
\frac{E_F}{n^3}=C_s\,\frac{16\alpha}{3\pi n^3} \mu_{\rm B}\mu_{\rm
nucl} R_\infty m_e^2\;,
\end{equation}
where we apply relativistic units in which $\hbar=c=1$,
$e^2/(4\pi)=\alpha$ is the fine structure constant, $m_e$ is the
electron mass, $R_\infty$ is the Rydberg constant, $\mu_{\rm B}$
is the Bohr magneton and $\mu_{\rm nucl}$ is the nuclear magnetic
moment. The normalization constant $C_s$ depends on the nuclear
spin. In particular, $C_s=1$ for the nuclear spin 1/2 (hydrogen,
helium-3 ion), while for the spin 1 (deuterium) an additional factor
$C_s=3/2$ appears.

A pseudovector particle, which interacts both with an electron and a
nucleus, would induce a spin-dependent interaction (cf.
contributions of the $Z$ boson \cite{weak} and $a_1$ meson
\cite{lbl} to the $1s$ HFS; see also \cite{prespin}). If such an
effect is present, the Coulomb exchange is modified by a
spin-dependent term
\begin{equation}\label{ss}
-\frac{Z\alpha}{r} \to -\frac{Z\left[\alpha+\alpha^{\prime\prime}
\bigl({\bf s}_e\cdot{\bf s}_{\rm N}\bigr)\,e^{-\lambda
r}\right]}{r}\;,
\end{equation}
where $Z$ is the nuclear change. Such a term is observable and may
be used to produce a constraint on $\alpha^{\prime\prime}(\lambda)$
while comparing an actual value of $D_{21}$ with theory.

In particular, in the limit
\[\lambda\ll Z\alpha m_e\sim
Z\cdot 3.5\;{\rm keV}\;,\]
energy of each HFS interval is shifted by
\begin{equation}
\Delta E_{\rm
hfs}(ns)=-C_s\,\frac{2}{n^2}\frac{\alpha^{\prime\prime}}{\alpha}\,(Z^2R_\infty)
\end{equation}
and the related contribution to the difference is
\begin{eqnarray}\label{d21cor}
\Delta D_{21}&=&-{2}C_s\frac{\alpha^{\prime\prime}}{\alpha}\,(Z^2R_\infty) \nonumber\\
&=&-0.9\times10^{18}\times C_s\,Z^2\times\alpha^{\prime\prime}\;{\rm
Hz}\;,
\end{eqnarray}
which should be compared with the difference between the related
experimental and theoretical values. The factor $C_s\,Z^2$ is unity
for hydrogen, 3/2 for deuterium, and 4 for the helium-3 ion.

\section{The constraint on the coupling constant of a pseudovector boson}

The present situation with experiments and theory of $D_{21}$ is
summarized in Table~\ref{T:exp}, which covers all available data on
determination of $D_{21}$in light two-body atoms. We also present
there a value of $\alpha^{\prime\prime}$ for an asymptotic region
$\lambda\ll 1\;$keV. The result is indeed consistent with zero,
since the theory and experiment are in perfect agreement.

\begin{table}[phtb]
\begin{tabular}{clll}
\hline
Atom & ~~~Experiment~ & ~~~~~~~Theory~~~ & ~~~~~~~~~~$\alpha^{\prime\prime}$  \\
 & ~~~~~~~~~[kHz] & ~~~~~~~~[kHz]  &    \\
 \hline
H & ~~~~~~48.923(54) & ~~~~~48.953(3)  & $~~\bigl(3.3\pm5.9\bigr)\times10^{-17}$ \\
D & ~~~~~~11.280(56) & ~~~~~11.3125(5) &  $~~\bigl(2.4\pm4.1\bigr)\times10^{-17}$ \\
$^3$He$^+$ &$-1189.979(71)$ &$-1190.08(15)$ & $\bigl(-2.8\pm4.6\bigr)\times10^{-17}$\\
\hline
\end{tabular}
\caption{Comparison of experiment and theory for the $D_{21}$ value
in light hydrogen-like atoms.  A negative sign for the HFS
difference for $^3$He$^+$ ion reflects the fact that the nuclear
magnetic moment is negative, i.e., in contrast to other nuclei in
the Table, its direction is antiparallel to the nuclear spin. The
constraint on $\alpha^{\prime\prime}$ is related to $\lambda\ll
1\;$keV. The confidence level of the constraint corresponds to one
standard deviation. \label{T:exp}}
\end{table}

If we consider $\alpha^{\prime\prime}$ as a certain universal
constant, an average value over the constraints in Table~\ref{T:exp}
is found as
\begin{equation}
\alpha^{\prime\prime}_{\rm
av}=\bigl(0.7\pm2.7\bigr)\times10^{-17}\;.
\end{equation}

To consider a constraint on a heavier intermediate particle, we have
to calculate the contribution of the Yukawa correction in (\ref{ss})
to the $D_{21}$ difference. As a result, the correction
(\ref{d21cor}) should include an additional factor ${\cal
F}_{12}(\lambda/(Z\alpha m_e))$ and the constraint takes the
form
\begin{equation}\label{a:profiled}
\alpha^{\prime\prime}(\lambda)=\frac{\alpha^{\prime\prime}_0}{{\cal
F}_{12}\bigl(\lambda/(Z\alpha m_e)\bigr)}\;,
\end{equation}
where $\alpha^{\prime\prime}_0$ is a constraint for
$\lambda/(Z\alpha m_e)\ll1$, listed in Table~\ref{T:exp} and the
profile function
\begin{eqnarray}
{\cal F}_{12}(x)&=& 4\left[\left(\frac{1}{1 +x}\right)^2-2
\left(\frac{1}{1+x}\right)^3+\frac{3}{2}
\left(\frac{1}{1+x}\right)^4\right]\nonumber\\
&&-\left(\frac{2}{2+x}\right)^2\nonumber
\end{eqnarray}
satisfies the condition ${\cal F}_{12}(x\to0)\to1$.

\begin{figure}[thbp]
\begin{center}
\resizebox{0.95\columnwidth}{!}{\includegraphics{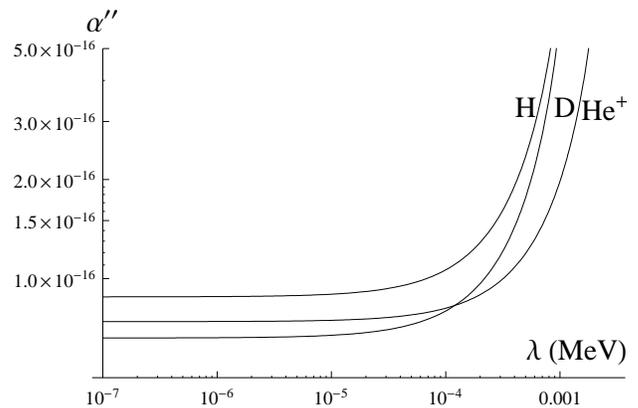}}
\end{center}
\caption{Constraints on a pseudovector intermediate boson from
$D_{21}$ in hydrogen, deuterium and helium-3 ion. The lines present
an upper bound for $\vert\alpha^{\prime\prime}\vert$. The
confidence level corresponds to one standard deviation.}
\label{f:const}       
\end{figure}

The related constraints extended to higher $\lambda$ are presented
in Fig.~\ref{f:const} \cite{prl}, however it has sharp $\lambda$
dependence and is not efficient above a few-keV level.

\section{Comparison
to other HFS constraints on pseudovector boson}

Because of low efficiency of the constraints in
Eq.~(\ref{a:profiled}) above the keV region, we have to combine the
results of this paper with constraint derived previously
\cite{prespin} from the data on the $1s$ HFS interval. Those
constraints are weaker in the keV range but they are more suitable
for extension to higher masses.

\begin{figure}[thbp]
\begin{center}
\resizebox{0.95\columnwidth}{!}{\includegraphics{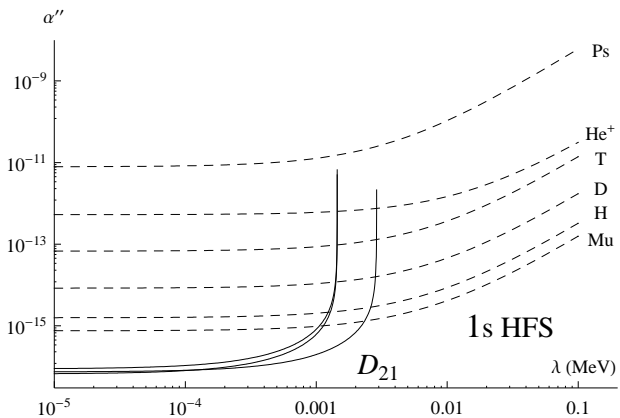}}
\end{center}
\caption{Constraints on a pseudovector intermediate boson from the
HFS study. The lines present the upper bound for
$\vert\alpha^{\prime\prime}\vert$ from data on $D_{21}$ (solid
lines; see Fig.~\ref{f:const} for detail) and the $1s$ HFS interval
(dashed lines) in various two-body atoms. The $1s$ results are from
\cite{prespin}. confidence level corresponds to one standard
deviation.}
\label{f:total}       
\end{figure}

The overall constraint \cite{prl} from a study of the hyperfine
intervals is summarized in Fig~\ref{f:total}. Three low lines are
from $D_{21}$ (cf. Fig.~\ref{f:const}) and the related constraints
are much stronger in the one-keV region and below, However, the
lines related to the $1s$ HFS interval \cite{prespin} produce
stronger constraints for above a few keV.

That is expectable. In the case of the Yukawa radius, longer than
atomic distances, the $D_{21}$ constraints gain in accuracy because
of the cancelation of the nuclear contributions which have large
uncertainties. (The same mechanism turns the $D_{21}$ difference
into a powerful tool to test bound state QED \cite{my_rep}.)
However, once the radius is shorter than atomic distances, the
Yukawa contribution becomes proportional to $|\Psi_{ns}(0)|^2$ and
it is canceled out almost completely. Technically, that shows up as
a special behavior of the function ${\cal F}_{12}(x)\propto x^{-4}$
at $x\to \infty$, while the related behavior for the $1s$
contribution \cite{prespin}
\begin{equation}
{\cal F}_{1}(x)=\left(\frac{2}{2+x}\right)^2\;,
\end{equation}
which in particular determines $\lambda$ dependence of the $1s$
constraints in Fig.~\ref{f:total}, is $\propto x^{-2}$. That makes
the $D_{21}$ difference insensitive to shorter-distance Yukawa
spin-spin interactions.

\begin{figure}[thbp]
\begin{center}
\resizebox{0.7\columnwidth}{!}{\includegraphics{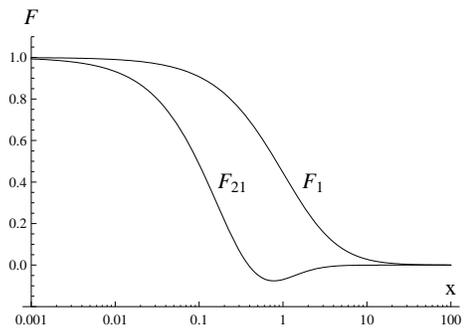}}
\end{center}
\caption{The profile functions giving the upper bound for
$\vert\alpha^{\prime\prime}\vert$ from data on $D_{21}$ and the $1s$
HFS interval (in various two-body atoms. The $1s$ results are from
\cite{prespin}.%
}
\label{f:profile}       
\end{figure}

For illustration, we present both profile functions in
Fig.~\ref{f:profile}. Both are equal to unity for low $\lambda$ and
that is the area, where the constraints are the strongest. At large
$\lambda$, both functions decrease to zero, which means that the
Yukawa correction vanishes. However, as we mentioned, the behavior
at high $\lambda$ is different, which produces a different
sensitivity for the high $\lambda$ region. The results are obtained
within a non-relativistic approximation. Taking into account
relativistic effects does not change sharp-edge behavior of ${\cal
F}_{12}$.

Thus, it is really fruitful to combine HFS constraints obtained by
both methods: the $D_{21}$ study for a longer wing of $\lambda$ and
the $1s$ HFS tests for the shorter one as summarized in
Fig~\ref{f:total}. The constraints derived are complementary to
various high-energy physics constraints reviewed in \cite{pdg}.

To conclude, we remind that in particle physics the vertex for an
interaction of a vector particle with a fermion is
$-ig_V\gamma_\mu$, while for the pseudovector it is
$-ig_A\gamma_5\gamma_\mu$. That means that the long-range
interaction for particles $x$ and $y$ mediated by a pseudovector
boson is of the form
\[
\frac{\alpha_A(xy) \bigl({\mbox{\boldmath$\sigma$}}_x\cdot
{\mbox{\boldmath$\sigma$}}_y\bigr)}{r} \;,
\]
where $\alpha_A(xy)=g_A(x)g_A(y)/(4\pi)$. Comparing with substitute
(\ref{ss}), where the spin-dependent coupling constant
$\alpha^{\prime\prime}$ is introduced, we note that
$\alpha_A=\alpha^{\prime\prime}/4$ (since ${\bf
s}_x={\mbox{\boldmath$\sigma$}}_x/2$). That is rather the constant
$\alpha_A$ that is the properly normalized coupling constant.

\begin{figure}[thbp]
\begin{center}
\resizebox{1.0\columnwidth}{!}{\includegraphics{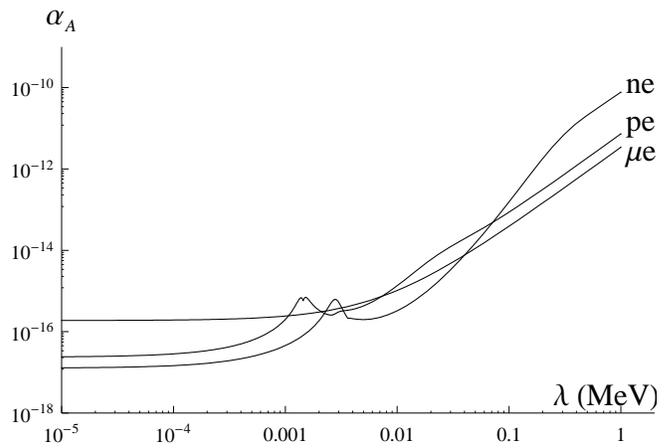}}
\end{center}
\caption{Constraints on a pseudovector intermediate boson. The lines
present the upper bound for the coupling constant
$\vert\alpha_A(xy)\vert$ for $xy=pe, ne, \mu e$ from data on HFS
intervals in various two-body atoms. The confidence level
corresponds to one standard deviation.}
\label{f:pnmu}       
\end{figure}

We summarize in Fig.~\ref{f:pnmu} the constraints on $\alpha_A(xe)$
for proton, neutron and muon (i.e. for $x=p, n, \mu$), where we have
taken into account all results derived in \cite{prespin} and in this
paper. To separate proton and neutron contributions we assume that
nuclear binding effects can be neglected and thus for the deuteron
we find
\[
\alpha_A({de})=\frac{ \alpha_A({pe}) + \alpha_A({ne})}{2}\;,
\]
while the helion constant is assumed to be equal to a free neutron
value ($\alpha_A({he})=\alpha_A({ne})$). Indeed, the binding effect
could add a certain additional uncertainty, which is to be
estimated. We do not think that would change the general situation.

\section*{Acknowledgments}

This work was supported in part by RFBR (grants \#\# 08-02-91969 \&
08-02-13516) and DFG (grant GZ 436 RUS 113/769/0-3). The author is
grateful to Andrej Afanasev, Dmitry Toporkov, Simon Eidelman, Eugene
Korzinin and Maxim Pospelov for useful and stimulating discussions.

\appendix

\section{Summary on experimental data on the $1s$ and $2s$ HFS intervals in light
two-body atoms\label{s:a1}}

The experimental results on the metastable $2s$ state are available
only for three hydrogen-like atoms, namely, for hydrogen, deuterium
and helium-3 ion. Only a few measurements have been performed for
over fifty years since fifties when the first results on the $2s$
HFS interval in hydrogen \cite{exph2sc} and deuterium \cite{expd2sa}
atoms and helium-3 ion \cite{exphe2sa} were obtained. We summarize
in Table~\ref{t:exp2} all obtained results.

\begin{table}[phtb]
\begin{tabular}{clc}
\hline
~~~~~~~~Atom~~~~~~~~ & ~~~~~~~~$E_{\rm HFS} ({\rm exp})$ ~~~~~~~~~~~~& ~~Refs.~~\\
 & ~~~~~~~~~~~~[kHz] &  \\
 \hline
Hydrogen & 177\,556.8343(67)&\cite{exph2s}\\
 & 177\,556.860(16)&\cite{exph2sa}\\
 & ~177\,556.785(29)&\cite{exph2sb}\\
 & ~177\,556.860(50)&\cite{exph2sc}\\
Deuterium & ~40\,924.454(7)&\cite{expd2s}\\
 & ~40\,924.439(20)&\cite{expd2sa}\\
$^3$He$^+$ ion & ~$- 1083\,354.980\,7(88)$&\cite{exphe2s}\\
 & ~$- 1083\,354.99(20)$&\cite{exphe2sa}\\
\hline
\end{tabular}
\caption{All results on the $2s$ HFS interval in light hydrogen-like
atoms obtained up to now.  A negative sign for the $^3$He$^+$ ion
reflects the fact that the nuclear magnetic moment is negative and
thus its direction is antiparallel to the nuclear spin.
\label{t:exp2}}
\end{table}

Since only these three atoms are important for calculations of a
specific difference of the HFS intervals in the $1s$ and $2s$
states, we collect in Table~\ref{t:exp1} the experimental results on
the $1s$ HFS interval for involved atoms.

\begin{table}[phtb]
\vspace{10pt}
\begin{tabular}{clc}
\hline
 ~~~~~~~~Atom~~~~~~~~ & ~~~~$E_{\rm hfs} ({\rm exp})$~~~~ & ~~Ref.~~\\
 & ~~~~~~~~[kHz] &  \\
 \hline
Hydrogen& 1\,420\,405.751\,768(1)&\cite{exph1}\\
Deuterium & ~~327\,384.352\,522(2) &\cite{expd1}\\
$^3$He$^+$ ion & - 8\,665\,649.867(10)&\cite{exphe1}\\
\hline
\end{tabular}
\caption{The most accurate results for the $1s$ HFS interval in
those light hydrogen-like atoms, for which the results on the $2s$
HFS interval are available.\label{t:exp1}}
\end{table}

The results on the difference $D_{21}$, based on the most accurate
experimental results, are present in Table~\ref{T:exp} of the paper.

\section{Summary on theory of the $D_{21}$ difference
in light two-body atoms\label{s:a2}}

A detailed review on theory of the $D_{21}$ difference in hydrogen,
deuterium and helium-3 ion can be found in \cite{d21th,my_rep}. The
results are summarized in Table~\ref{T:d21th}. `QED3' and `QED4'
stands for pure QED corrections in units of the Fermi energy $E_F$,
defined in (\ref{ef}).

There are three small parameters in QED theory: $\alpha$ stands for
QED loops and is for the QED perturbation effects, $Z\alpha$ is for
the Coulomb strength and describes binding effects, while the mass
ratio $m/M$ (electron-to-nucleus) is for the recoil effects in
two-body atoms. Theoretical evaluations have a certain history,
being started in \cite{mittleman,zwanziger,sternheim}, shortly after
the first results on the $2s$ HFS interval were achieved
\cite{expd2sa,exph2sc,exphe2sa}.

The QED3 term involves various combinations of these three
parameters up to the third-order, which were mainly calculated long
time ago. A more recent development was due to the fourth-order
contributions (QED4), which include the fourth-order contributions
and, due to higher-order nuclear effects.

\begin{table}[phtb]
\begin{tabular}{clll}
\hline
~~Contribution~~  & Hydrogen & Deuterium & $^3$He$^+$ ion\\
to $D_{21}$& ~~~~[kHz] & ~~~~[kHz] & ~~~[kHz] \\
 \hline
$D_{21}({\rm QED3})$ & 48.937 &  11.305\,6 & -1\,189.253\\
$D_{21}({\rm QED4})$ & {0.018(5)} & {0.004\,4(10)}  &-1.13(14)\\
$D_{21}({\rm Nucl})$ & {-0.002} & {0.002\,6(2)} & 0.307(35)\\
\hline
$D_{21}({\rm total})$ & 48.953(5) &  11.312\,5(10) & -1\,190.08(15)  \\
\hline
\end{tabular}
\caption{Theory of the specific difference $D_{21}$ in light
hydrogen-like atoms \cite{d21th1}. The numerical results are
presented for the related frequency $D_{21}/h$. QED3 and QED4 stands
for the third- and fourth-order QED corrections in units of the Fermi
energy $E_F$ (see \cite{d21th,my_rep} for detail). \label{T:d21th}}
\end{table}

As we mentioned above, there is a substantial cancelation of the
nuclear-structure contribution in difference $D_{21}$. The leading
term, which takes into account nuclear charge and magnetic moment
distribution, cancels completely. However, certain higher-order
nuclear-effect contributions survive the cancelations and they are
denoted as `Nucl'. Those higher-order terms were found in
\cite{d21th1}.

For QED3 terms and for higher-order nuclear effects we follow
\cite{my_rep}, while for the QED4 terms we apply results of
\cite{d21th1}, a recent correction in which follows a reexamination
of the former QED4 calculation in \cite{d21th} and numerical
medium-$Z$ calculation of one-loop effects in \cite{oneloop} (cf.
\cite{oneloop1}).

\end{document}